\documentclass[
 reprint,
]{revtex4-1}

\usepackage{graphicx}
\usepackage{dcolumn}
\usepackage{mathrsfs}
\usepackage{amsmath}
\usepackage{bm}
\newtheorem{theorem}{Theorem}

\begin{document}

\preprint{APS/123-QED}

\title{Using NonBacktracking Expansion to Analyze $k$-core Pruning Process}

\author{Rui-jie Wu}
\author{Yi-Xiu Kong}
\author{Gui-yuan Shi}
\email{guiyuan.shi@unifr.ch}
\author{Yi-Cheng Zhang}
 
\affiliation{Department of Physics, University of Fribourg, Fribourg 1700, Switzerland}

\date{\today}

\begin{abstract}

We induce the NonBacktracking Expansion Branch method to analyze the $k$-core pruning process on the monopartite graph $G$ which does not contain any self-loop or multi-edge. Different from the traditional  approaches like the generating functions or the degree distribution evolution equations which are mathematically difficult to solve, this method provides a simple and intuitive solution of the $k$-core pruning process. Besides, this method can be naturally extended to study the $k$-core pruning process on correlated networks, which is among the few attempts to analytically solve the problem.

\end{abstract}

\pacs{Valid PACS appear here}
\maketitle

\section{Introduction}

$k$-core decomposition is a widely used method for identifying the center of a large network. It is a pruning process in which nodes with degrees less than $k$ are recursively removed. It has many applications across different fields like biology, informatics, economy, network science, etc\cite{seidman1983network,bader2003automated,lahav2016k,garas2010worldwide,kitsak2010identification,carmi2007model}. and attracts many theoretical studies. One of the most important question is the size of the final $k$-core, it has been studied analytically by previous researches\cite{fernholz2004cores, dorogovtsev2006k}. They also find that there exists a critical phenomenon in $k$-core decomposition. The detailed process of the $k$-core has become increasingly interesting for many researchers. Baxter et al. \cite{baxter2015critical} proposed a theoretical framework consists of 4 equations to analyze $k$-core pruning process, and obtain the recursive relationship of the degree distribution between step $n$ and step $n + 1$.
Shi et al. \cite{shi2018analytical} further simplified this calculation to an univariable iteration process by inducing an auxiliary series $y_n$, and the asymptotic results when $n\rightarrow \infty$ is consistent with previous researches of final state. 
However in general, the theoretical framework in previous study\cite{baxter2015critical} are difficult to solve analytically. Here we use the NonBacktracking Expansion Branch method which greatly simplified the calculation process. The results are consistent with the previous results \cite{shi2018analytical}. In addition, our theoretical results can be easily extended to correlated networks, which has always been difficult to solve analytically.

\section{Method}

We are concerned about the $k$-core pruning process on the monopartite graph $G$ which does not contain any self-loop or multi-edge. In this paper we will show the NonBacktracking Expansion Branch can be used to solve the problem for both uncorrelated networks and correlated networks. 

Firstly, we give the definition of the NonBacktracking Expansion Branch(NBEB) of the graph which is introduced in ref.\cite{weiss2001optimality,timar2017nonbacktracking}. But here we build the NBEB by using the concept of the stubs $(e,V)$ introduced by M. Newman\cite{newman2018networks}. Note that it is equivalent to the branch $B(\infty,i\gets j)$ used in ref.\cite{timar2017nonbacktracking}. For a given node $V$ in a graph $G$ and assume the degree of $V$ is $j$, let $\{V_1,V_2, \dots V_j\}$ to be the neighbors of $V$, and $\{e_1,e_2, \dots e_j\}$ are the edges connecting them to $V$. Then we define the neighbor stubs set $S(V)=\{(e_1,V_1),(e_2,V_2),\dots,(e_j,V_j)\}$ is the set of stubs connected to $V$. If $j\geq 1$, for any stub $(e,V)$, we denote edge $e$ of node $V$ as $e_j$, and define excess neighbor stubs set $S(e,V)=\{(e_1,V_1),(e_2,V_2),\dots,(e_{j-1},V_{j-1})\}$ is a set of stubs connected to $V$ except the stub that connects to node $V$ through $e$. Starting from a randomly chosen stub $(e,V)$, now we can define a NBEB which is a tree-like structure, the root stub $(e,V)$ can be regarded as the first layer of the tree, then the second layer of the NBEB consists of the child nodes of the root, which is the set of stubs $S(e,V)$. 


Consequently for each stub $(e^*,V^*)$ in the $n^{th}$ layer, we can find the child nodes of $(e^*,V^*)$ which are the stubs in $S(e^*,V^*)$, and all these child stubs are the $(n+1)^{th}$ layer of th NBEB.

\begin{figure*}[htb]
\centering
\includegraphics[width=13cm]{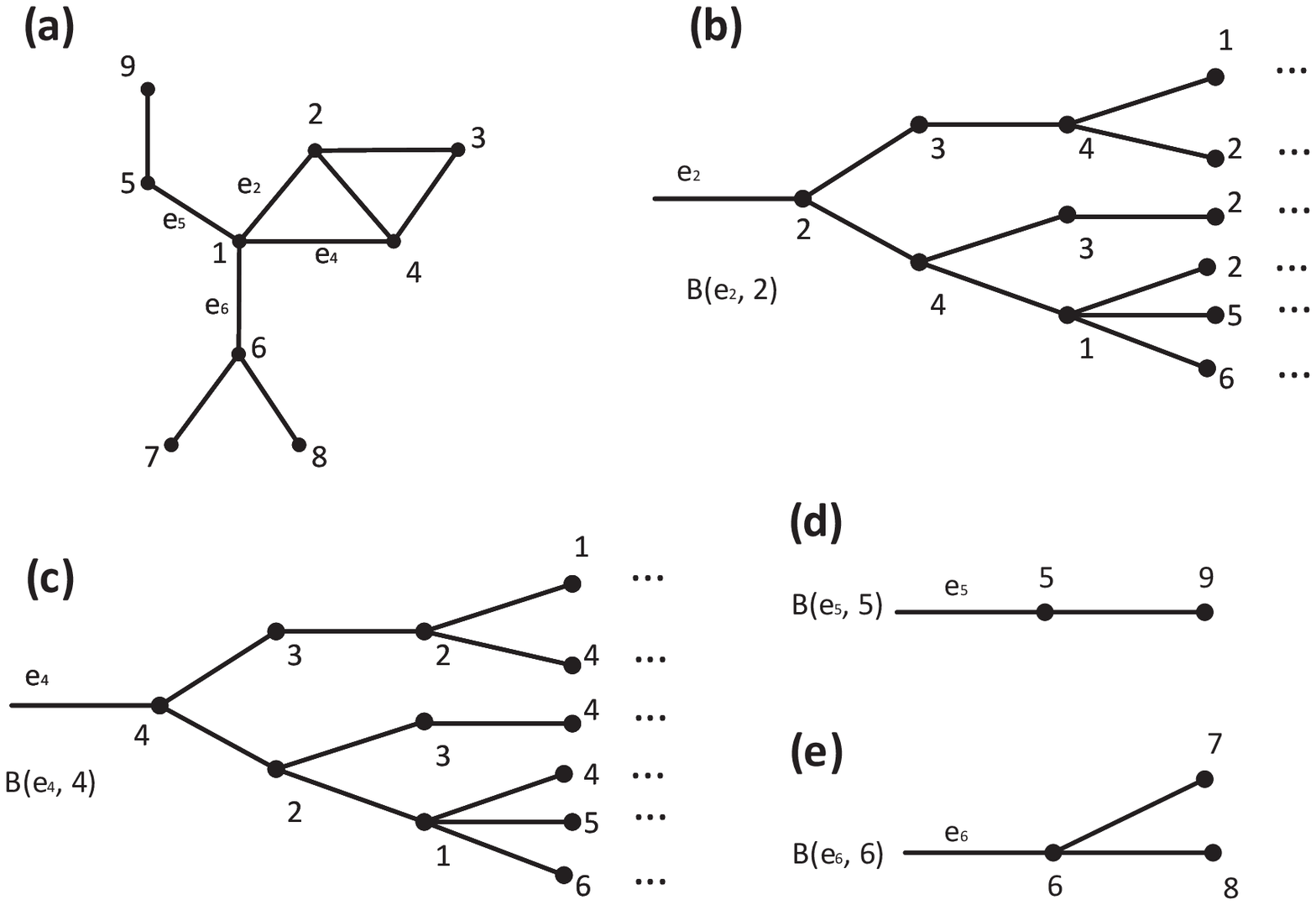}
\caption{A simple example of NBEB method on a simple network. \textbf{a)} A simple network consists of 9 nodes and 10 edges. \textbf{b-e)} the NBEBs $B(e_2,2)$, $B(e_4,4)$, $B(e_5,5)$, $B(e_6,6)$ respectively. Note that some of NBEBs are infinite so that we only show the first few layers of the branch. 
}
\label{tree2}
\end{figure*}

We can continue this process so that we obtain a NBEB of the stub $(e,V)$ which is denoted by $B(e,V)$. $B(e,V)$ can be either infinite or finite depending on the structure of the network. It is easy to know for any given finite graph $G$ with $M$ edges, we can have $2M$ NBEBs. In Fig. \ref{tree2}\textbf{a} we have a simple network, Fig. \ref{tree2}\textbf{b} shows a simple example of $B(e_2,2)$ based on the network in Fig. \ref{tree2}\textbf{a}. The neighbor stubs set of node $1$ $S(1)=\{(e_2,2),(e_4,4),(e_5,5),(e_6,6)\}$, the excess neighbor stubs set of $(e_2,1)$ is $S(e_2,1)=\{(e_4,4),(e_5,5),(e_6,6)\}$. The first few layers of the NBEBs of the target node $1$ are also shown in Fig. \ref{tree2}\textbf{b-e}.


\section{Analysis of $k$-core pruning process}

As we previously introduced, $k$-core pruning process is a process that we recursively remove the nodes whose degrees are less than $k$. In the following we discuss how the NBEB method can be utilized to solve the $k$-core pruning process analytically. 
For a given positive integer $k$, we can find a set of NBEBs $Y_n$($n$ is a positive integer) which meet the following condition: there exists a subbranch of the NBEB which contains the root stub, and the amount of child nodes of each node in the first $n$ layers of this subbranch are no less than $k-1$. Generally we denote $Y_0$ to be the set of all the NBEBs. Obviously, $Y_0\supset Y_1 \supset \dots \supset Y_n \supset Y_{n+1} \supset \dots$. For convenience, we denote $S_{NBEB}(V)$ to be the set of NBEBs of all the stubs in $S(V)$, and $S_{NBEB}(e,V)$ to be the set of NBEBs of all the stubs in $S(e,V)$. In the simple example in Fig. \ref{tree2}, $S_{NBEB}(1)$ consists of $B(e_2,2)$, $B(e_4,4)$, $B(e_5,5)$, $B(e_6,6)$, and $S_{NBEB}(e_2,1)$ consists of $B(e_4,4)$, $B(e_5,5)$, $B(e_6,6)$.
According to the definition of $Y_n$ it is easy to prove:

\begin{theorem}
The NBEB $B(e,V)$ belongs to the set $Y_n$ if and only if among the NBEBs in $S_{NBEB}(e,V)$, at least $k-1$ NBEBs belong to $Y_{n-1}$.  
\end{theorem}

Besides, let $S_n$ be the subgraph of the remaining nodes after $n^{th}$ pruning, and the following theorem can be established,

\begin{theorem}
Denote by $V$ a node in the graph $G$,  $V\in S_n$ if and only if among the NBEBs in $S_{NBEB}(V)$ at least $k$ NBEBs belong to $Y_{n-1}$.
\end{theorem}

The proof of Theorem 2 is given in Appendix. Besides, to further clarify the meaning of Theorem 2, we present Fig. \ref{tree2} as an example of how to analyze $k$-core pruning process with NBEB method, the detailed process is given in Appendix.

\subsection{Uncorrelated Networks}
In this part we analyze the $k$-core pruning process on large($N\rightarrow \infty$, $N$ is the size of the network) uncorrelated networks using our NBEB approach. 

We assume the probability that a randomly chosen NBEB belongs to $Y_n$ is $y_n$. Obviously we have $y_0=1$, from Theorem 1 we know $y_n$ should meet the following recurrence relation:

\begin{align*}
y_{n}&=\sum_{j=k-1}^{\infty}\sum_{m=k-1}^{j}q_j{j \choose m}y_{n-1}^m(1-y_{n-1})^{j-m} \\
             &=\sum_{m=k-1}^{\infty}\frac{y_{n-1}^m}{m!}\sum_{j=m}^{\infty}\frac{j!}{(j-m)!}q_j(1-y_{n-1})^{j-m}\\
             &=\sum_{m=k-1}^{\infty}\frac{y_{n-1}^m}{m!}*G_1^{(m)}(1-y_{n-1})\\
              &=1-\sum_{m=0}^{k-2}\frac{y_{n-1}^m}{m!}*G_1^{(m)}(1-y_{n-1})
\end{align*}
The above formula satisfies the same recursive relationship with the $y_n$ sequence defined in our previous study\cite{shi2018analytical}. In the previous paper, we induced the auxiliary series $y_n$ to facilitate the induction of the evolution of the $k$-core pruning process, but $y_n$ was only regarded as a mathematical trick to solve the problem. Here we find that $y_n$ has a practical probabilistic meaning that it represents the probability that a randomly chosen NBEB belongs to $Y_n$.

Let us set $s_n$ as the portion of the remaining nodes $S_n$ relative to the total number of all nodes $N$. Theorem 2 gives:
\begin{align*}
s_{n}&=\sum_{j=k}^{\infty}\sum_{m=k}^{j}p_j{j \choose m}y_{n-1}^m(1-y_{n-1})^{j-m} \\
             &=\sum_{m=k}^{\infty}\frac{y_{n-1}^m}{m!}\sum_{j=m}^{\infty}\frac{j!}{(j-m)!}p_j(1-y_{n-1})^{j-m}\\
             &=\sum_{m=k}^{\infty}\frac{y_{n-1}^m}{m!}*G_0^{(m)}(1-y_{n-1})\\
              &=1-\sum_{m=0}^{k-1}\frac{y_{n-1}^m}{m!}*G_0^{(m)}(1-y_{n-1})
\end{align*}

This is consistent with our results in the previous paper. the final size of $k$-core can be obtained easily if taking the limit of $n\rightarrow \infty$. 

In the previous paper\cite{shi2018analytical}, we defined $v_n$ which is the probability that, if a randomly selected stub $(e,V)$ from the remaining network after the $n^{th}$ pruning(the network is defined as $N_n$), the excess degree of $V$ in the $N_n$ is greater than or equal to $k-1$. Previously we obtained the relationship $ v_n = y_ {n + 1} / y_n $ by mathematical induction. Here we show the relationship can be easily obtained by NBEB method, and also can be given a probabilistic meaning. 

For an edge $e$, both of its endpoints $V$ and $V^*$ survive after $n^{th}$ pruning, means that the edge $e$ will be retained in the $N_n$. If the excess degree of $V$ in the $N_n$ is greater than or equal to $k-1$, which suggests $V$ will survive after $(n+1)^{th}$ pruning but for $V^*$ it may or may not be retained after the $(n+1)^{th}$ pruning. 

So $v_n$ can also be interpreted equivalently to: for an edge $e$, and its two endpoints $V$ and $V^*$ in the original network, the probability that $V$ will survive in $N_{n+1}$ given the condition that both $V$ and $V^*$ survived after $n^{th}$ pruning.

Based on the NBEB method we introduced above, suppose we have a random edge $e$ in the original network, denote its two endpoints by $V$ and $V^*$ respectively. If $e$ survives after the $n^{th}$ pruning, then we have this condition if and only if both $B(e,V)$ and $B(e,V^*)$ belong to $Y_n$, the probability of this occurrence is equal to $y_n^2$. If the endpoint $V^*$ survives at $n^{th}$ pruning, endpoint $V$ survives at $(n+1)^{th}$ pruning, then this happens if and only if $B(e,V^*)$ belongs to $Y_n$, and $B(e,V) $ belongs to $Y_{n+1}$, this probability is equal to $y_n*y_{n+1}$. Obviously the second event is included in the first one. Now consider the conditional probability of occurrence of the second event given that the first event has occurred is: $y_n*y_{n+1}/y_n^2=y_{n+1}/y_n$, which is exactly the definition of $v_n$ shown above.

\subsection{Correlated Networks}

In the following we discuss the $k$-core pruning process on large correlated networks using the above method, we take the representation of M. Newman\cite{newman2002assortative}, using $e_{jl}$ to represent the probability that an edge has an endpoint whose excess degree is $j$ and the other endpoint has an excess degree of $l$ , then it is easy to see $e_{jl}=e_{lj}$, and:

\begin{equation*}
\sum_{l=0}^{\infty}e_{jl}=q_j
\end{equation*}

If a network is uncorrelated, then we have:  $e_{jl}=q_j*q_l$.

Now consider the $k$-core of the correlated network. Consider in a NBEB $B(e,V)$ that the excess degree of other endpoint of edge $e$ is $l$, the probability of $B(e,V)\in Y_n$ is denoted by $y_{n,l}$. Obviously when $q_l=0$, the conditional probability $y_{n,l}$ does not exist, we can set $y_{n,l}=1$. From Theorem 1, it satisfies the following recursion: 

\begin{equation*}
q_l*y_{n,l}=\sum_{j=k-1}^{\infty}e_{lj}\sum_{m=k-1}^{j}{j \choose m}y_{n-1,j}^m*(1-y_{n-1,j})^{j-m}
\end{equation*}

Therefore, Theorem 2 gives that $s_n$ satisfies the following relationship:
\begin{equation*}
s_{n}=\sum_{j=k}^{\infty}p_{j}\sum_{m=k}^{j}{j \choose m}y_{n-1,j-1}^m*(1-y_{n-1,j-1})^{j-m}
\end{equation*}   

The final size of the $k$-core can be obtained by taking the limit $n\rightarrow \infty$.
Consider an extreme case, let $e_{jl}=q_j*q_l$, which means the network is an uncorrelated network, it is easy to know that if $q_l\ne 0$, $q_j\ne 0$, $y_{n,l}=y_{n, j}$, and we can denote it by $y_n$, then we have the same conclusion as the uncorrelated case discussed above. Another extreme case is that when the network is completely correlated\cite{newman2002assortative}, i.e., $e_{lj}=q_j\delta_{jl}$, we can easily find: when $j\geq k$, $y_{n,j}=1$, $s_n=\sum_{j=k}^{\infty}p_j$. In this case only two nodes with the same degree are possible to be connected, so all of nodes whose degree are less than $k$ will be deleted, and all of the nodes whose degrees are greater than or equal to $k$ are retained.

\section{Conclusion}

In this paper we find that the NonBacktracking Expansion Branch(NBEB) method can be used to analyze the $k$-core pruning process. The NBEB method offers a more intuitive perspective to analyze the $k$-core pruning process, and it greatly simplifies the calculation. The results are consistent with the previous analytical results, and also provide a practical probabilistic meaning of the auxiliary series we previously induced to facilitate the induction. Besides, we find that this method can be easily extended to deal with the correlated networks, thus provides an unprecedented theoretical tool to understand the $k$-core organizations in real-world networks.

\section{Appendix}

\noindent
\textbf{Proof of Theorem 2}:
\\

We use mathematical induction to prove the theorem. It is obvious that the theorem holds for $n=1$. Now we prove that if the theorem is true for $n-1$,  the theorem can be established for $n$.

Firstly we prove the sufficiency, that is, when at least $k$ NBEBs in $S_{NBEB}(V)$ belong to $Y_{n-1}$, we have $V\in S_n$. Since $Y_{n-1}\subset Y_{n-2}$, obviously at least $k$ NBEBs in $S_{NBEB}(V)$ belong to $Y_{n-2}$, so from the inductive hypothesis $V\in S_{n-1}$, on the other hand, we suppose that the NBEBs in the subset $\{B(e_{i_1}, V_{i_1}), B(e_{i_2}, V_{i_2}), \dots, B(e_{i_m}, V_{i_m})\}\subset S_{NBEB}(V)$ belong to $Y_{n-1}$, where $m\geq k$, so for any $1\leq r \leq m$, in $S_{NBEB}(e_{i_r}, V_{i_r})$, At least $k-1$ NBEBs belong to $Y_{n-2}$ since $B(e_{i_r}, V_{i_r})\in Y_{n-1}$, furthermore, $B(e_{i_r},V)\in Y_{n} \subset Y_{n-2}$ since except $B(e_{i_r},V_{i_r})$ there are still at least $k-1$ NBEBs in $S_{NBEB}(e_{i_r}, V)$ belong to $Y_{n-1}$. Therefore, for each $V_{e_r}$, at least $k$ of the NBEBs in $S_{NBEB}(V_{i_r})$ belong to $Y_{n-2}$. The induction hypothesis gives $V_{e_r}\in S_{n-1}$, as a result, in the $(n-1)^{th}$ pruning, at least $k$ neighbors of the node $V$ are retained in the $(n-1)^{th}$ pruning. We can then conclude $V$ is still retained in the $n^{th}$ pruning.

Next we prove the necessity. we will prove that when at most $k-1$ NBEBs in $S_{NBEB}(V)$ belong to $Y_{n-1}$, there must be $V\notin S_{n}$. Because for a NBEBs $B(e_r, V_r)$ that does not belong to $Y_{n-1}$,  There are at most $k-2$ NBEBs in $S_{NBEB}(e_r,V_r)$ belong to $Y_{n-2}$, which is known from the definition of $Y_n$. Then we know there exist at most $k-1$ NBEBs in $S_{NBEB}(V_r)$ that belong to $Y_{n-2}$. Since the induction hypothesis gives $V_r\notin S_{n-1}$, therefore, among all the neighbors of node $V$, at most $k-1$ neighbors are retained in the $(n-1)^{th}$ pruning. So either $V$ has been deleted in the $(n-1)^{th}$ pruning or even before, or it is retained in $N_{n-1}$, but removed in $n^{th}$ pruning because its remaining neighbors are less than $k$ after the $(n-1)^{th}$ pruning, so $V\notin S_{n}$.

At this point, the sufficiency and necessity are proved, and the theorem 2 can be established. \\

\noindent
\textbf{Detailed process of NBEB analysis}
\\

We use the network shown in Fig. \ref{tree2} as an example. For $2$-core pruning process, we are to judge whether the target node $1$ will be retained in the $n^{th}$ pruning of $2$-core. Node $1$ has 4 NBEBs as shown in Fig.\ref{tree2}\textbf{b-e}. According to Theorem 2, Node 1 will be retained in the $n^{th}$ pruning of $2$-core if and only if at least $2$ NBEBs in $S_{NBEB}(1)$ belong to $Y_{n-1}$. Obviously all 4 branches in $S_{NBEB}(1)$ belong to $Y_0$ and $Y_1$. Although $B(e_5,5)$ and $B(e_6,6)$ do not belong to $Y_2$, $B(e_2,2)$ and $B(e_4,4)$ belong to $Y_\infty$ which ensures $2$ branches in $S_{NBEB}(1)$ belong to $Y_{n-1}$ for any step $n$. Thus we can conclude node $1$ keeps staying in the $2$-core.  

Similarly,  Node 1 will be retained in the $n^{th}$ pruning of $3$-core if and only if at least 3 NBEBs in $S_{NBEB}(1)$ belong to $Y_{n-1}$. Obviously all 4 branches in $S_{NBEB}(1)$ belong to $Y_0$, among them only $B(e_2,2)$, $B(e_4,4)$ and $B(e_6,6)$ belong to $Y_1$, none of the 4 branches belong to $Y_2$.  so we know that node $1$ will be kept at the first two steps of pruning and deleted at the $3^{rd}$ pruning.

\nocite{*}

\bibliography{bibliography}

\begin{thebibliography}{14}%
\makeatletter
\providecommand \@ifxundefined [1]{%
 \@ifx{#1\undefined}
}%
\providecommand \@ifnum [1]{%
 \ifnum #1\expandafter \@firstoftwo
 \else \expandafter \@secondoftwo
 \fi
}%
\providecommand \@ifx [1]{%
 \ifx #1\expandafter \@firstoftwo
 \else \expandafter \@secondoftwo
 \fi
}%
\providecommand \natexlab [1]{#1}%
\providecommand \enquote  [1]{``#1''}%
\providecommand \bibnamefont  [1]{#1}%
\providecommand \bibfnamefont [1]{#1}%
\providecommand \citenamefont [1]{#1}%
\providecommand \href@noop [0]{\@secondoftwo}%
\providecommand \href [0]{\begingroup \@sanitize@url \@href}%
\providecommand \@href[1]{\@@startlink{#1}\@@href}%
\providecommand \@@href[1]{\endgroup#1\@@endlink}%
\providecommand \@sanitize@url [0]{\catcode `\\12\catcode `\$12\catcode
  `\&12\catcode `\#12\catcode `\^12\catcode `\_12\catcode `\%12\relax}%
\providecommand \@@startlink[1]{}%
\providecommand \@@endlink[0]{}%
\providecommand \url  [0]{\begingroup\@sanitize@url \@url }%
\providecommand \@url [1]{\endgroup\@href {#1}{\urlprefix }}%
\providecommand \urlprefix  [0]{URL }%
\providecommand \Eprint [0]{\href }%
\providecommand \doibase [0]{http://dx.doi.org/}%
\providecommand \selectlanguage [0]{\@gobble}%
\providecommand \bibinfo  [0]{\@secondoftwo}%
\providecommand \bibfield  [0]{\@secondoftwo}%
\providecommand \translation [1]{[#1]}%
\providecommand \BibitemOpen [0]{}%
\providecommand \bibitemStop [0]{}%
\providecommand \bibitemNoStop [0]{.\EOS\space}%
\providecommand \EOS [0]{\spacefactor3000\relax}%
\providecommand \BibitemShut  [1]{\csname bibitem#1\endcsname}%
\let\auto@bib@innerbib\@empty
\bibitem [{\citenamefont {Seidman}(1983)}]{seidman1983network}%
  \BibitemOpen
  \bibfield  {author} {\bibinfo {author} {\bibfnamefont {S.~B.}\ \bibnamefont
  {Seidman}},\ }\href@noop {} {\bibfield  {journal} {\bibinfo  {journal}
  {Social networks}\ }\textbf {\bibinfo {volume} {5}},\ \bibinfo {pages} {269}
  (\bibinfo {year} {1983})}\BibitemShut {NoStop}%
\bibitem [{\citenamefont {Bader}\ and\ \citenamefont
  {Hogue}(2003)}]{bader2003automated}%
  \BibitemOpen
  \bibfield  {author} {\bibinfo {author} {\bibfnamefont {G.~D.}\ \bibnamefont
  {Bader}}\ and\ \bibinfo {author} {\bibfnamefont {C.~W.}\ \bibnamefont
  {Hogue}},\ }\href@noop {} {\bibfield  {journal} {\bibinfo  {journal} {BMC
  bioinformatics}\ }\textbf {\bibinfo {volume} {4}},\ \bibinfo {pages} {2}
  (\bibinfo {year} {2003})}\BibitemShut {NoStop}%
\bibitem [{\citenamefont {Lahav}\ \emph {et~al.}(2016)\citenamefont {Lahav},
  \citenamefont {Ksherim}, \citenamefont {Ben-Simon}, \citenamefont
  {Maron-Katz}, \citenamefont {Cohen},\ and\ \citenamefont
  {Havlin}}]{lahav2016k}%
  \BibitemOpen
  \bibfield  {author} {\bibinfo {author} {\bibfnamefont {N.}~\bibnamefont
  {Lahav}}, \bibinfo {author} {\bibfnamefont {B.}~\bibnamefont {Ksherim}},
  \bibinfo {author} {\bibfnamefont {E.}~\bibnamefont {Ben-Simon}}, \bibinfo
  {author} {\bibfnamefont {A.}~\bibnamefont {Maron-Katz}}, \bibinfo {author}
  {\bibfnamefont {R.}~\bibnamefont {Cohen}}, \ and\ \bibinfo {author}
  {\bibfnamefont {S.}~\bibnamefont {Havlin}},\ }\href@noop {} {\bibfield
  {journal} {\bibinfo  {journal} {New Journal of Physics}\ }\textbf {\bibinfo
  {volume} {18}},\ \bibinfo {pages} {083013} (\bibinfo {year}
  {2016})}\BibitemShut {NoStop}%
\bibitem [{\citenamefont {Garas}\ \emph {et~al.}(2010)\citenamefont {Garas},
  \citenamefont {Argyrakis}, \citenamefont {Rozenblat}, \citenamefont
  {Tomassini},\ and\ \citenamefont {Havlin}}]{garas2010worldwide}%
  \BibitemOpen
  \bibfield  {author} {\bibinfo {author} {\bibfnamefont {A.}~\bibnamefont
  {Garas}}, \bibinfo {author} {\bibfnamefont {P.}~\bibnamefont {Argyrakis}},
  \bibinfo {author} {\bibfnamefont {C.}~\bibnamefont {Rozenblat}}, \bibinfo
  {author} {\bibfnamefont {M.}~\bibnamefont {Tomassini}}, \ and\ \bibinfo
  {author} {\bibfnamefont {S.}~\bibnamefont {Havlin}},\ }\href@noop {}
  {\bibfield  {journal} {\bibinfo  {journal} {New journal of Physics}\ }\textbf
  {\bibinfo {volume} {12}},\ \bibinfo {pages} {113043} (\bibinfo {year}
  {2010})}\BibitemShut {NoStop}%
\bibitem [{\citenamefont {Kitsak}\ \emph {et~al.}(2010)\citenamefont {Kitsak},
  \citenamefont {Gallos}, \citenamefont {Havlin}, \citenamefont {Liljeros},
  \citenamefont {Muchnik}, \citenamefont {Stanley},\ and\ \citenamefont
  {Makse}}]{kitsak2010identification}%
  \BibitemOpen
  \bibfield  {author} {\bibinfo {author} {\bibfnamefont {M.}~\bibnamefont
  {Kitsak}}, \bibinfo {author} {\bibfnamefont {L.~K.}\ \bibnamefont {Gallos}},
  \bibinfo {author} {\bibfnamefont {S.}~\bibnamefont {Havlin}}, \bibinfo
  {author} {\bibfnamefont {F.}~\bibnamefont {Liljeros}}, \bibinfo {author}
  {\bibfnamefont {L.}~\bibnamefont {Muchnik}}, \bibinfo {author} {\bibfnamefont
  {H.~E.}\ \bibnamefont {Stanley}}, \ and\ \bibinfo {author} {\bibfnamefont
  {H.~A.}\ \bibnamefont {Makse}},\ }\href@noop {} {\bibfield  {journal}
  {\bibinfo  {journal} {Nature physics}\ }\textbf {\bibinfo {volume} {6}},\
  \bibinfo {pages} {888} (\bibinfo {year} {2010})}\BibitemShut {NoStop}%
\bibitem [{\citenamefont {Carmi}\ \emph {et~al.}(2007)\citenamefont {Carmi},
  \citenamefont {Havlin}, \citenamefont {Kirkpatrick}, \citenamefont
  {Shavitt},\ and\ \citenamefont {Shir}}]{carmi2007model}%
  \BibitemOpen
  \bibfield  {author} {\bibinfo {author} {\bibfnamefont {S.}~\bibnamefont
  {Carmi}}, \bibinfo {author} {\bibfnamefont {S.}~\bibnamefont {Havlin}},
  \bibinfo {author} {\bibfnamefont {S.}~\bibnamefont {Kirkpatrick}}, \bibinfo
  {author} {\bibfnamefont {Y.}~\bibnamefont {Shavitt}}, \ and\ \bibinfo
  {author} {\bibfnamefont {E.}~\bibnamefont {Shir}},\ }\href@noop {} {\bibfield
   {journal} {\bibinfo  {journal} {Proceedings of the National Academy of
  Sciences}\ }\textbf {\bibinfo {volume} {104}},\ \bibinfo {pages} {11150}
  (\bibinfo {year} {2007})}\BibitemShut {NoStop}%
\bibitem [{\citenamefont {Fernholz}\ and\ \citenamefont
  {Ramachandran}(2004)}]{fernholz2004cores}%
  \BibitemOpen
  \bibfield  {author} {\bibinfo {author} {\bibfnamefont {D.}~\bibnamefont
  {Fernholz}}\ and\ \bibinfo {author} {\bibfnamefont {V.}~\bibnamefont
  {Ramachandran}},\ }\href@noop {} {\bibfield  {journal} {\bibinfo  {journal}
  {The University of Texas at Austin, Department of Computer Sciences,
  technical report TR-04-13}\ } (\bibinfo {year} {2004})}\BibitemShut {NoStop}%
\bibitem [{\citenamefont {Dorogovtsev}\ \emph {et~al.}(2006)\citenamefont
  {Dorogovtsev}, \citenamefont {Goltsev},\ and\ \citenamefont
  {Mendes}}]{dorogovtsev2006k}%
  \BibitemOpen
  \bibfield  {author} {\bibinfo {author} {\bibfnamefont {S.~N.}\ \bibnamefont
  {Dorogovtsev}}, \bibinfo {author} {\bibfnamefont {A.~V.}\ \bibnamefont
  {Goltsev}}, \ and\ \bibinfo {author} {\bibfnamefont {J.~F.~F.}\ \bibnamefont
  {Mendes}},\ }\href@noop {} {\bibfield  {journal} {\bibinfo  {journal}
  {Physical review letters}\ }\textbf {\bibinfo {volume} {96}},\ \bibinfo
  {pages} {040601} (\bibinfo {year} {2006})}\BibitemShut {NoStop}%
\bibitem [{\citenamefont {Baxter}\ \emph {et~al.}(2015)\citenamefont {Baxter},
  \citenamefont {Dorogovtsev}, \citenamefont {Lee}, \citenamefont {Mendes},\
  and\ \citenamefont {Goltsev}}]{baxter2015critical}%
  \BibitemOpen
  \bibfield  {author} {\bibinfo {author} {\bibfnamefont {G.}~\bibnamefont
  {Baxter}}, \bibinfo {author} {\bibfnamefont {S.}~\bibnamefont {Dorogovtsev}},
  \bibinfo {author} {\bibfnamefont {K.-E.}\ \bibnamefont {Lee}}, \bibinfo
  {author} {\bibfnamefont {J.}~\bibnamefont {Mendes}}, \ and\ \bibinfo {author}
  {\bibfnamefont {A.}~\bibnamefont {Goltsev}},\ }\href@noop {} {\bibfield
  {journal} {\bibinfo  {journal} {Physical Review X}\ }\textbf {\bibinfo
  {volume} {5}},\ \bibinfo {pages} {031017} (\bibinfo {year}
  {2015})}\BibitemShut {NoStop}%
\bibitem [{\citenamefont {Shi}\ \emph {et~al.}(2018)\citenamefont {Shi},
  \citenamefont {Wu}, \citenamefont {Kong}, \citenamefont {Stanley},\ and\
  \citenamefont {Zhang}}]{shi2018analytical}%
  \BibitemOpen
  \bibfield  {author} {\bibinfo {author} {\bibfnamefont {G.-Y.}\ \bibnamefont
  {Shi}}, \bibinfo {author} {\bibfnamefont {R.-J.}\ \bibnamefont {Wu}},
  \bibinfo {author} {\bibfnamefont {Y.-X.}\ \bibnamefont {Kong}}, \bibinfo
  {author} {\bibfnamefont {H.~E.}\ \bibnamefont {Stanley}}, \ and\ \bibinfo
  {author} {\bibfnamefont {Y.-C.}\ \bibnamefont {Zhang}},\ }\href@noop {}
  {\bibfield  {journal} {\bibinfo  {journal} {arXiv preprint arXiv:1810.08936}\
  } (\bibinfo {year} {2018})}\BibitemShut {NoStop}%
\bibitem [{\citenamefont {Weiss}\ and\ \citenamefont
  {Freeman}(2001)}]{weiss2001optimality}%
  \BibitemOpen
  \bibfield  {author} {\bibinfo {author} {\bibfnamefont {Y.}~\bibnamefont
  {Weiss}}\ and\ \bibinfo {author} {\bibfnamefont {W.~T.}\ \bibnamefont
  {Freeman}},\ }\href@noop {} {\bibfield  {journal} {\bibinfo  {journal} {IEEE
  Transactions on Information Theory}\ }\textbf {\bibinfo {volume} {47}},\
  \bibinfo {pages} {736} (\bibinfo {year} {2001})}\BibitemShut {NoStop}%
\bibitem [{\citenamefont {Tim{\'a}r}\ \emph {et~al.}(2017)\citenamefont
  {Tim{\'a}r}, \citenamefont {da~Costa}, \citenamefont {Dorogovtsev},\ and\
  \citenamefont {Mendes}}]{timar2017nonbacktracking}%
  \BibitemOpen
  \bibfield  {author} {\bibinfo {author} {\bibfnamefont {G.}~\bibnamefont
  {Tim{\'a}r}}, \bibinfo {author} {\bibfnamefont {R.~A.}\ \bibnamefont
  {da~Costa}}, \bibinfo {author} {\bibfnamefont {S.~N.}\ \bibnamefont
  {Dorogovtsev}}, \ and\ \bibinfo {author} {\bibfnamefont {J.~F.}\ \bibnamefont
  {Mendes}},\ }\href@noop {} {\bibfield  {journal} {\bibinfo  {journal}
  {Physical Review E}\ }\textbf {\bibinfo {volume} {95}},\ \bibinfo {pages}
  {042322} (\bibinfo {year} {2017})}\BibitemShut {NoStop}%
\bibitem [{\citenamefont {Newman}(2018)}]{newman2018networks}%
  \BibitemOpen
  \bibfield  {author} {\bibinfo {author} {\bibfnamefont {M.}~\bibnamefont
  {Newman}},\ }\href@noop {} {\emph {\bibinfo {title} {Networks}}}\ (\bibinfo
  {publisher} {Oxford university press},\ \bibinfo {year} {2018})\BibitemShut
  {NoStop}%
\bibitem [{\citenamefont {Newman}(2002)}]{newman2002assortative}%
  \BibitemOpen
  \bibfield  {author} {\bibinfo {author} {\bibfnamefont {M.~E.}\ \bibnamefont
  {Newman}},\ }\href@noop {} {\bibfield  {journal} {\bibinfo  {journal}
  {Physical review letters}\ }\textbf {\bibinfo {volume} {89}},\ \bibinfo
  {pages} {208701} (\bibinfo {year} {2002})}\BibitemShut {NoStop}%
\end{thebibliography}%

\end{document}